\documentstyle[twoside,11pt]{article}\textheight 200 mm\textwidth 140 mm

\newcommand{\be}{\begin{eqnarray}}
\newcommand{\ee}{\end{eqnarray}}
\newcommand{\np}{\newpage}
\newcommand{\jt}{\tilde{J}}

\newcommand{\pr}{\partial}
\newcommand{\ti}{\tilde}
\newcommand{\ng}{|0\rangle_{gh}}
\newcommand{\hs}{\hspace}
\newcommand{\vs}{\vspace}
\newcommand{\al}{\alpha}
\newcommand{\de}{\delta}

\newcommand{\ka}{\kappa}
\newcommand{\ind}{\indent}
\newcommand{\cSG}{{\cal S}_G}
\newcommand{\cSH}{{\cal S}_H}
\newcommand{\cStH}{{\cal S}_{\widetilde{H}}}

\newcommand{\ket}[1]{|#1\rangle}

\begin{document}

\begin{flushright}
G\"oteborg ITP 93-01 \\
hepth 9305174 \\
\vs{1mm}
\noindent May 1993\\
\vs{15mm}
\end{flushright}

\begin{center}
{\LARGE The BRST formulation of G/H WZNW models.}\\
\vs{20mm}
{\Large Stephen Hwang\footnote{tfesh@fy.chalmers.se}
 and Henric Rhedin\footnote{hr@fy.chalmers.se}} \vs{10mm}

Institute of Theoretical Physics \\
University of G\"oteborg \\
and \\
Chalmers University of Technology \\
\vs{10mm}

{\bf Abstract }  \\

\end{center}

We consider a BRST approach to G/H coset WZNW models, {\it i.e.} a formulation
 in which the coset is defined by a BRST condition. We will give the precise
ingrediences needed for this formulation. Then we will prove the equivalence
of this approach to the conventional coset formulation by
solving the the BRST cohomology. This will reveal a remarkable connection
 between integrable representations and a class of non-integrable
representations for negative levels. The latter representations are also
 connected to string theories based on non-compact WZNW models.
The partition functions of G/H cosets are also considered. The BRST approach
 enables a covariant construction of these, which does not rely on the
decomposition of G as $G/H\times H$. We show that for the well-studied
 examples of $SU(2)_k \times SU(2)_1/SU(2)_{k+1}$ and $SU(2)_k/U(1)$, we
 exactly reproduce the previously known results.

\np
\pagestyle{myheadings}
\setcounter{page}{2}
\section{Introduction}
\setcounter{equation}{0}
\markboth{\underline{\hspace{89mm} \  \bf{Section} \thesection \ \ Introductio
n }}
{\underline{\bf{Section} \thesection \ \ Introduction \hspace{89mm}
 \  }}
In the study of two-dimensional conformal field theories the affine Lie
algebras have played an essential r\^ole. One of the most important
 constructions, in this connection, is the so-called coset construction
of which examples first appeared in \cite{H} and then were constructed in
general by Goddard, Kent and Olive (GKO) \cite{GKO}. The
 corresponding models, known as the gauged Wess-Zumino-Novikov-Witten (WZNW)
 models, have a symmetry algebra which are of the affine Lie type \cite{W1} and
 appear to describe most of the known conformal field theories. It is,
 therefore, of importance to give a fundamental formulation of these models,
 enabling a consistent construction of tree and loop
amplitudes.

A key to such a fundamental formulation was first given by Karabali and
 Schnitzer \cite{KS}, who showed that the gauged $G/H$ WZNW models resulted
in an action which was BRST invariant. This action contained three different
sectors. Apart from the original $G$ WZNW model, it was described by an
auxiliary non-unitary $H$ WZNW model and a Faddeev-Popov ghost sector. The
 nilpotent BRST symmetry results from a conversion
of the original constraints of the gauged WZNW model, which are of second
 class classically \cite{Bo}, to new constraints which are classically of
 first class.
One may also view the BRST symmetry as a consequence of a change of
variables \cite{Ba}.

In an operator formulation, the BRST invariance implies that one should
require that physical states of the gauged WZNW model should be BRST
 invariant. One must then analyze this equation and solve the resulting BRST
 cohomology.
This will give the possible states of the theory. In the case of $H$ being
Abelian, it was proved \cite{KS} that the BRST approach was equivalent, up
to a degeneracy due to the ghost zero modes, to the usual coset conditions.
It was, furthermore, shown that for arbitrary $G$ and $H$, the resulting
energy-momentum tensor differed from the GKO one by a BRST exact term. It is
the purpose of this paper to complete the work of Karabali and Schnitzer. We
 will first give the exact ingrediences needed to treat general $G$ and $H$.
 In particular, we will specify the representations of the auxiliary $H$ WZNW
 model. The choice of representations will be one of the most essential parts
 in our work. We will then solve the BRST cohomology. The result of this
 analysis will show the equivalence of the BRST formulation and conventional
 coset
 construction. As a consequence of this analysis, we will discover an
 intriguing connection between
the integrable representations of $H$ and a class of non-integrable
 representations of
$H$ with negative levels.

A BRST formulation of gauged WZNW models will be of
importance in a further development of these theories. For example, we will
 study the partition function of the $G/H$ models in our formulation. We will
 derive the general contributions of the auxiliary and ghost sectors.
 Combined with the
contribution of the $G$ WZNW model, which are given by the Kac-Weyl formula for
simple $G$, they will give the resulting partition function for the $G/H$
model.
This will give a
covariant construction, which does not rely on a decomposition of $G$ as
$G/H\times H$. We study in detail two simple examples, $SU(2)_k \times
 SU(2)_1/SU(2)_{k+1}$ and $SU(2)_k/U(1)$, for which we exactly reproduce the
 previously known results. The partition functions for arbitrary
$G/H$ models have also been treated using a path-integral approach \cite{GK}.
Their derivation does not, however, seem to respect the BRST symmetry.

We will also briefly discuss correlation functions. Using a decoupling
 theorem, originally proved in
connection with a string theory based on the $SU(1,1)$ WZNW model \cite{HwRo},
 which is a modification of the original theorem due to Gepner and Witten
 \cite{GW},
we can prove that our choice of representations for the auxiliary theory is
 consistent in correlation functions. A remaining problem, which is not
 addressed, is to determine fusion rules for the auxiliary theory.

The representations of the auxiliary non-unitary WZNW theory for the simplest
 case, $SU(2)$, are the discrete infinite dimensional non-unitary
representations and they correspond to the unitary discrete ones of $SU(1,1)$.
The range of highest weights $j$, $k/2<j<0$, $(k<0)$ are the ones which have
unitary physical states in the so-called $SU(1,1)$ string \cite{Hwang}, as
well as for the coset $SU(1,1)/U(1)$ \cite{DLP} and will also contain the range
of admissible representations in our BRST formulation. This connection is quite
intriguing and might prove to be a way to understand the models based on
non-compact groups. In particular, it may explain why these models seem to
be so much simpler for integer values of $k$ \cite{HHRS}. Furthermore, using
known fusion rules for the compact cosets one may deduce the fusion rules
of the non-compact theory. The latter are at present not known.

The BRST formulation has attracted some attention in treating topological
$G/G$ theories \cite{AGSYS}. There one does not impose any
restrictions on the representations of the auxiliary $G$ theory. This will
lead to a much more complicated structure of the BRST cohomology. To analyze
 the BRST cohomology one has used a free field realization. This
realization, together with another nil-potent operator, introduced by Felder
for minimal models \cite{F} and later generalized to $SU(2)_k$ \cite{BeFe},
has provided an effective tool for analyzing the cohomology. This has
 {\it e.g.}
been demonstrated in connection with $c\leq 1$ matter coupled to 2D gravity
\cite{BMP1}. The problem in treating arbitrary groups $G$ is that, in general,
 the corresponding Felder reduction has not been rigourously proven. By
 assuming that
a reduction exists, some general results on the cohomology \cite{BMP2} and the
partition functions \cite{BMP3} can be derived. The methods that we will use
 in this
paper
 are not based on a free field realization. Instead they rely on a
 formalism developed in \cite{HM}. We have some general
results, using our techniques, for the cohomology of general $G$ and more
 general
representations of the auxiliary $H$ theory. We intend to present these
 results in a
future publication.

\section{The gauged WZNW model}
\setcounter{equation}{0}
\markboth{\underline{\hspace{61mm} \  \bf{Section} \thesection \ \ The gauged
 WZNW model }}
{\underline{\bf{Section} \thesection \ \ The gauged WZNW model \hspace{60mm}
 \  }}

\ \\
\ind We consider a general WZNW model defined on a Riemann surface
 ${\cal M}$ with fields $g$ taking values in a compact Lie group $G$.
 The action is \cite{W1} \cite{GW} \cite{KZ}.
\be
S_{\bf k}(g) =\frac{1}{16\pi}\int_{\cal M} d^2\xi
 Tr(\pr_{\mu}g^{-1}\pr^{\mu}g)+
\frac{1}{24\pi}\int_B d^3x\epsilon^{\alpha\beta\gamma}Tr(g^{-1}\pr _{\alpha}g
g^{-1}\pr _{\beta}g
g^{-1}\pr_{\gamma}g)\label{action}.
\ee
Here we assume that $g$ is well-defined on a three-dimensional manifold $B$,
 which has ${\cal M}$ as boundary. ${\bf k}$ refers to levels $k_i$ according
 to the decomposition of $G$, $G=G_1\times G_2\times \ldots$, where $G_i$
 are simple or belong to the center of $G$. We will assume that all $k_i$ are
 positive and integers and that we only have integrable representations. From
 here on we will, for simplicity, consider a simple group with a level $k$.
 Our analysis may be extended to the general case.

This action is invariant under the transformations \cite{W1}.
\be
g(\xi)\longrightarrow\Omega(z)g(\xi)\bar\Omega^{-1}(\bar{z})\label{gtrans}
\ee
$\Omega$, and $\bar{\Omega}$ are G-valued matrices analytically depending on
$z=\xi_1+i\xi_2$ $\bar{z}=\xi_1-i\xi_2$ respectively.
The symmetry (\ref{gtrans}) implies an infinite number of conserved
currents.
\be
\pr_{\bar{z}}J=0 \hs{25mm} \pr_{z}\bar{J}=0
\ee
These currents $J=J^At_A$ and $\bar{J}=\bar{J^A}t_A$, with $t_A$ antihermitean
 matrices representing the Lie
algebra $g$ of the group $G$ \footnote{We use here the same notation $g$ for
 the algebra as for the fields in the WZNW theory. It should be clear from the
 context what is meant.}, satisfy the affine Lie algebra $\hat{g}$
\be
[J^A_m,J^B_n]=if^{AB}_{~~C}J^C_{m+n}+\frac{k}{2}m\delta_{m+n}g^{AB},\label{km}
\ee
with a corresponding algebra for $\bar{J}^A_m$. $f^{AB}_{~~C}$ are structure
 constants of $g$ and $g^{AB}$ a non-degenerate metric on $G$.

In order to gauge an anomaly-free vector subgroup $H$ of the global $G\times
 G$  symmetry one introduces the gaugefields $A$, which belong to the adjoint
 representation of $H$. We denote the level of $H$ by $k_{H}$ and we assume
 again, for simplicity, $H$ to be simple. The
corresponding action is then in a light-cone decomposition
\be
S_{k}(g,A)=S_{k}(g)+\frac{1}{4\pi}\int \hs{-1mm}d^2\xi
Tr(A_+\pr_-gg^{-1}\hs{-1mm}-A_-g^{-1}\pr_+g+
A_+gA_-g^{-1}-\hs{-1mm}A_-A_+)\label{actioni}
\ee
By integrating out the gauge fields and using the Polyakov-Wiegmann identity,
 the partition function may be written in the form \cite{GK},\cite{KPSY}
\be
Z=&\int&[dg][d\tilde{h}][db_+][db_-][dc_+][dc_-]exp[-kS_{k}(g)]
exp[-(-k_{H}-2c_H)S_{- k_H-2{c}_H}(\tilde{h})]\nonumber\\
&\times& exp[-Tr\int d^2\xi (b_+\pr_-c_++b_-\pr_+c_-)]\label{part}
\ee
The partition function for the gauged $G/H$ model factorizes, therefore, into
  three different sectors: The original WZNW theory with a group $G$ and level
 $k$, an auxiliary WZNW theory with a group $H$ and level $-k_{H}-2c_{H}$,
 where $c_{H}$ is the second Casimir of the adjoint representation of $H$,
 and a ghost sector. The total action in (\ref{part}) is invariant under BRST
transformations \cite{KS}\cite{Ba}
\be
\delta_Bg&=&c_-g-gc_+ \nonumber \\
\delta_Bh&=&c_-h-hc_+ \nonumber \\
\delta_Bc_{\pm}&=&-\frac{1}{2}\{c_{\pm},c_{\pm}\} \nonumber \\
\delta_Bb_+&=&-\frac{k}{4\pi}g^{-1}\pr_+g+\frac{k_H+c_H}{4\pi}h^{-1}\pr_
+h-\{b_+,c_+\} \nonumber \\
\delta_Bb_-&=&\frac{k}{4\pi}\pr_-gg^{-1}-\frac{k_H+c_H}{4\pi}\pr_-hh^{-1}
-\{b_-,c_-\}     \label{brst}
\ee
In writing the partition function (\ref{part}) we have not been precise in
 defining the auxiliary WZNW theory. One must specify what representations
occur for this sector. Since its level $-k_{H}-2c_{H}$ is negative it is not,
 in general, unitary. As a consequence the choice of representations is not
 restricted in the same way as in the conventional unitary WZNW theory. A
possible choice is the principal series of continuous representations
 \cite{GK}. We will return to this issue in analyzing the BRST cohomology.
 In particular, we will see that choosing only
the principal series is not consistent with the BRST symmetry. \vspace{5mm}

The BRST symmetry found in the path-integral approach above is, however, also
 natural from an operator point of view. The action (\ref{actioni}) implies
 classically that $J^a(z)\approx 0$ and $\bar{J}^a(\bar{z})\approx 0$, where
 $a$
 take values in $h$, the Lie algebra of $H$\footnote{Our conventions are such
 that indices $A,B,\ldots$ take
values in $g$ and $a,b,\ldots$ take values in $h$.}. These constraints are of
 second class at the classical level \cite{Bo}. In quantizing the theory
 canonically one should impose, in some way, such constraints on physical
states. This is also implied in the Goddard-Kent-Olive (GKO) construction
 of coset theories \cite{GKO}, where these currents commute with
 energy-momentum tensor of the coset. It is natural to impose on physical
 states
\be
J^a_n|{\rm phys}\rangle=0, \label{coset}
\ee
where $n>0$ or, $n=0$  and $a$ being a positive root. We will refer to
 (\ref{coset}) as the conventional coset conditions.

A more fundamental way of imposing
constraints in a quantum theory is by using the BRST symmetry. In this case we
 have an obstruction
due to the second class nature of the constraints, implying a BRST charge
 which is not nilpotent. To overcome this difficulty one introduces a new
 auxiliary set of variables, in order to convert the constraints into first
 class ones. A systematic construction was first discussed in \cite{BF}.
The auxiliary theory must give rise
to a change in the original constraint generators $J^a\rightarrow J^a+\jt^a$
 such that
the corresponding BRST charge is nilpotent at the quantum level. We are,
 therefore, led to the following BRST charge for the chiral part
\be
Q=\oint\frac{dz}{2i\pi}\left[:c_a(z)(J^a(z)+\jt^a(z)):-\frac{i}{2}f^{ad}_{\ \
 e}
:c_a(z)c_d(z)b^e(z):\right]\label{brstq}
\ee
The modes of the currents $J^a(z)$ and $\jt^a(z)$ satisfy an $\hat{h}$
 affine Lie algebra with levels $k_H$ and $k_{\widetilde{H}}$. We will assume
 that $J^a(z)$ will depend linearly on the currents of $G$. The ghosts
 $c_a(z)$ and $b^a(z)$ are conformal fields of dimension zero and one,
 respectively. They satisfy an operator product expansion $c_a(z)b^b(w)
=\frac{\delta_{a}^{\ b}}{z-w}$. It is straightforward
to check the nilpotency of the BRST charge. It holds provided we take
 $k_{\widetilde{H}}=-k_H-2c_H$, which is the value found in the path-integral
 approach. Therefore, in a canonical quantization of a gauged WZNW theory,
 the auxiliary theory of level $-k_H-2c_H$ arises from the requirement of a
 nilpotent BRST charge. The physical states are then found as solutions of
\be
Q|{\rm phys}\rangle=0\label{brsteq}
\ee
The purpose of this paper is to investigate what the implications of this
 condition are. In particular, if the naive conditions (\ref{coset}) are
 consistent with the BRST approach.

The nilpotency of the BRST charge implies that the current
\be
J^{tot,a}(z)=[Q,b^a(z)]=J^a(z)+\jt^a(z)+J^{gh,a}(z)
\ee
satisfies an $\hat{h}$ affine Lie algebra with vanishing central term. Here
 $J^{gh,a}(z)=if^{ad}_{\ \ e}:b^{e}(z)c_d(z):$, which satisfies an $\hat{h}$
 affine Lie algebra of level $2c_H$.

{}From the partition function (\ref{part}) one may deduce the holomorphic
 energy-momentum tensor
\be
T(z)&=&\frac{1}{k+c_G}:J_A(z)J^A(z):-\frac{1}{k+c_H}:\jt_a(z)\jt^a(z):-
:b^{a}(z)\pr_zc_a(z): \nonumber \\
&\equiv &T^G+\ti{T}^H+T^{gh} \hs{50mm}
\ee
which has a total conformal anomaly
\be
c^{tot}=\frac{kd_G}{k+c_G}+\frac{(-k-2c_H)d_H}{(-k-2c_H)+c_H}-2d_H=
\frac{kd_G}{k+c_G}-\frac{kd_H}{k+c_H}
\ee
We recognize this charge as being equivalent to the one from the
 energy-momentum tensor in the GKO construction \cite{GKO}, $T^{GKO}=T^G-T^H$.
 The connection between the two is even more clear if we write $T(z)
=T^{GKO}(z)+T^H(z)+T^{\widetilde{H}}(z)+T^{gh}(z)$, where \cite{KS}
\be
T^H(z)+T^{\widetilde{H}}(z)+T^{gh}(z)=\frac{1}{k+c_H}\left[Q,:b_a(z)
\left( J^a(z)-\jt^a(z)
\right):\right]\label{emq}
\ee
The fact these two energy-momentum tensors differ only by a BRST exact term
suggests that the conventional coset construction is at least contained in the
 BRST approach.
 For $H$ being Abelian it was proved \cite{KS} that the two constructions are
 equivalent {\it i.e.} that (\ref{brsteq}) and (\ref{coset}) yield the same
 spectrum of physical states\footnote{The BRST approach yields a two-fold
 degeneracy of physical states due to the doubling of ghost vacua.}. The
 situation for more general $H$ is, however, more complicated and this we
 will discuss in the following sections.

For future reference, let us introduce the Cartan-Weyl basis. In this basis
 the
affine Lie algebra reads
\be
[J_m^i,J_n^j]&=&\frac{k}{2} m\delta^{ij}\delta_{m,-n}
\nonumber \\
{}~[J_m^i, J_n^\alpha ]&=& \alpha ^iJ_{m+n}^\alpha
\nonumber\\
 ~[J_m^\al,J_n^\beta] &=& \left\{\begin{array}{ll} \epsilon(\al,\beta)J_{m+n}
^{\al+\beta}\hs{30mm} &
\mbox{if $\alpha +\beta$ is a root}
 \\
\frac{1}{\al^2}(\al_iJ^i_{m+n}+\frac{k}{2}m\de_{m,-n}) \hs{7mm} & \mbox{ if
 $\al=-\beta$ } \\
 0 \hs{48mm} & \mbox{otherwise.}\label{cwbasis}
\end{array}
\right.
\ee
Here $i$ label the Cartan subalgebra and $\alpha$ are the roots, which are
 normalized so that $\alpha^2=1$ for the long roots. The ghosts are labelled
 by $c^i_n$,$b^i_n$,$c^\alpha_n$ and $b^\alpha_n$ with the non-zero
 anti-commutation relations
$\{c^i_m,b^j_n\}=\de_{m+n,0}\de^{ij}$ and $\{c^\al_m,b^\beta_n\}=\de^{\al,
-\beta}\de_{m,-n}$
The BRST charge is then
\be
Q&=&\sum_{i,n}:c_{-n}^i(J_{n}^i+\jt_{n}^i):+\sum_{\al,n} :c_{-n}^{\alpha}
(J_{n}^{-\alpha}+\jt_{n}^{-\alpha}):\nonumber\\
&-&
\frac{1}{2}\sum_{\alpha\neq -\beta}\sum_{m,n\in\cal{Z}}\epsilon(\al,\beta)
:b_{-m-n}^{\alpha+\beta}c^{-\alpha}_mc^{-\beta}_n
-\frac{1}{2}\sum_{i,\alpha}\sum_{m,n\in\cal{Z}}\frac{1}{\alpha^2}\alpha^i
:b_{-m-n}^{i}c^{-\alpha}_mc^{\alpha}_n:\nonumber\\
&+&
\sum_{i,\alpha}\sum_{m,n\in\cal{Z}}\alpha^i:b_{-m-n}^{\alpha}c^{-\alpha}_m
c^{i}_n:\label{qcw}
\ee
Furthermore
\be
J^{tot,i}_m\equiv \{Q,b_n^i\}= J^i_m+\jt_m^i+\sum_{\alpha ,n}\alpha^i
:b^\alpha_{m-n} c^{-\alpha}_n: \label{jtot}
\ee
and
\be
L^{tot}_m&\equiv &L^H_m+\ti{L}_m^H+
L^{gh}_m =\{Q,\frac{1}{k+c_H}\sum_{n}(J_{m+n}^a-\jt_{m+n}^a)b_{-n ,a}\}
\label{ltot}
\ee
where
\be
L^H_m&=&\frac{1}{k+c_H}\left(\sum_{i,n}:J_n^iJ_{m-n}^i:+
 \sum_{\al,n}\alpha^2:J_n^{-\alpha}J_{m-n}^\alpha :\right)\nonumber\\
\ti{L}_m^H&=&-\frac{1}{k+c_H}\left(\sum_{i,n} \jt_n^i\jt_{m-n}^i:+
 \sum_{\al,n}\alpha^2:\jt_n^{-\alpha}\jt_{m-n}^\alpha :\right) \nonumber\\
L^{gh}_m&=&\sum_{i,n} n:b_{m-n}^ic_n^i:+\sum_{\al,n} n:b_{m-n}^{-\alpha}
 c_n^{\alpha}:\label{lm}
\ee

\section{The BRST cohomology}
\setcounter{equation}{0}
\markboth{\underline{\hspace{66mm} \  \bf{Section} \thesection \ \ The BRST
 cohomology }}
{\underline{\bf{Section} \thesection \ \ The BRST cohomology \hspace{66mm}
 \  }}

We will now analyze the BRST equation (\ref{brsteq}). First we specify more
 exactly the state space. It decomposes into a product of three different
 sectors. A general
state is of the form $|s\rangle=|s_G\rangle\times |s_{\tilde H}\rangle\times
 |s_{gh}\rangle$, where
\be
| s_G\rangle =\sum_R\prod_{A,n}J^A_{-n}|0;R\rangle,\hs{5mm}| s_{\widetilde{H}}
\rangle =\sum_{\tilde R}\prod_{a,n}\jt_{-n}^a| \tilde{0};\widetilde{R}\rangle,
 \hs{25mm} \label{state}
\ee
and $| s_{gh}\rangle$ is a sum of states of the form
\be
\prod_{a_1,n_1}\prod_{a_2,n_2}b_{-n_1}^{a_1}c_{-n_2}^{a_2}| 0\rangle_{gh}.
\ee
Here the state $|0\rangle_{gh}$ is the SL(2, R) invariant ghostvacuum, which
 by the requirement that it is annihilated by $L^{gh}_n$ n=0,$\pm1$ satisfies
\be
c^a_n\ng=0 \hs{10mm} n\geq1 \hs{25mm} b_{n}^a\ng=0 \hs{10mm} n\geq0.
\ee
The state $|0;R\rangle$ is a highest weight primary state with respect to
 currents of $\hat{g}$, which transforms in some representation $R$ of $g$.
 The corresponding product in (\ref{state}) is taken over $n>0$ or $n=0$ and
 $A$ being a negative root. Similarly, we
have the state $|\tilde{0};\widetilde{R}\rangle$ which is highest weight
 primary with respect to $\hat{\tilde{h}}$ and $n>0$. If there exists a
 highest weight state of the finite algebra then we include $n=0$ and
 $a\in\{\alpha^-\}$ ({\it i.e.} the set of negative roots). The primary states
 in eq.(\ref{state}) have both zero occupation level.

Among the states in eq.(\ref{state}) there will exist, in general, states
 which have vanishing inner product with all
states. These states are nullstates. The set of states $| s_G\rangle$ and $|
 s_{\widetilde{H}}\rangle$ in eq.(\ref{state}), which are not null, will be
 denoted
$\cSG$ and $\cStH$, respectively. The full set of states
 $\cSG\times\cStH\times{\cal S}_{gh}$ is denoted ${\cal S}$. The
 corresponding set of states for the subalgebra $\hat{h}$ of $\hat{g}$ will
 be denoted $\cSH$.   It will be assumed throughout that any non-zero state in
the space $\cSG$ has a positive definite inner product.

It is convenient to
rewrite the states in the $G$-sector to exhibit the explicit dependence on the
currents of $\hat{h}$. We rewrite eq.(\ref{state}) as
\be
| s_G\rangle=\sum_{\Phi}\prod_{a,n}J^a_{-n}|\Phi\rangle\label{decomp}.
\ee
for some set of states $|\Phi\rangle\in \{\Phi\}$. If all states in this set
 are primary
with respect to the currents of $\hat{h}$, then the state space of the
 $G$-sector will completely decompose into representations of $\hat{h}$.
 Symbolically we can write this as $G=G/H \times H$, where $G/H$ is the set
 of primaries with respect to $\hat{h}$. It will also imply a corresponding
 decomposition of the partition function. If, however, $\{\Phi\}$ is not
equivalent to the set of primary states, then this decomposition does not
 hold. One may establish the following criterion
 ({\it decomposition theorem})\vs{5mm}

\begin{quotation}{\it The decomposition eq.(\ref{decomp}) of an arbitrary
 state in $\cSG $ is possible,
for all states in $\{\Phi\}$ being primary w.r.t. the currents of $\hat{h}$,
if and only if all null-states w.r.t. $\hat{h}$ are also null w.r.t.
 $\hat{g}$.\vs{5mm}}
\end{quotation}

\noindent We will not give the proof of this theorem here, but refer to a
 forthcoming publication. The theorem above is quite general and may be
 stated for other symmetry algebras than affine Lie algebras. In the present
 case we will not need this general result. For the theories that we are
 considering, namely when we restrict
to compact $G$ and integrable representations, a theorem due to Kac and
Peterson \cite{KACPET} states that it is always possible to choose $\{\Phi\}$
 as the set
of primary states w.r.t. the currents of $\hat{h}$. As remarked above, this
implies a corresponding decomposition of the characters of $G$. They may be
 written in terms of characters of
$H$ and so called {\it branching functions}.
 \vs{5mm}

The state space in eq.(\ref{state}) decomposes completely in eigenstates of
 $L_0$ of each sector, respectively. We assume throughout that these
 eigenvalues are finite. In analyzing the BRST equation (\ref{brsteq}), one
 may restrict oneself to definite eigenvalues of $L_0^{tot}$, since this
 operator commutes with the BRST charge. One has
\be
L_0^{tot}|s\rangle=(\frac{1}{k+c_H}(C-\widetilde{C})+N_J+\ti{N}_{\jt}+N_{gh})|s
\rangle
\ee
where $C$ and $\ti{C}$ are quadratic Casimirs of the finite algebra spanned by
  $J^a_0$, and
$\jt^a_0$,
respectively. $N$ is the total mode-number for each sector defined as the sum
 of
the individual modes. By a standard argument, only the states with zero
 eigenvalue of $L_0^{tot}$ are non-trivial in the BRST cohomology. This
 implies that for positive values of $C-\widetilde{C}$ all BRST invariant
 states are BRST exact, since $N_J+\ti{N}_{\jt}+N_{gh}$ is a positive
quantity. In the $H$-sector the values of $C$ that may occur are non-negative,
 since only unitary finite
dimensional representations are possible. We can conclude, therefore, that in
 the $\widetilde{H}$-sector all representations for which $\widetilde{C}$ is
 strictly negative
will have only trivial solutions to the BRST equation. This is the case for
 the principal series of continuous representations. We can, consequently,
 restrict our attention to representations in the $\widetilde{H}$-sector for
 which the Casimir eigenvalues are non-negative.

Let the Cartan subalgebra of $h$ be denoted by $J^i_0$,  $\tilde{J}^i_0$ and
 $J^{gh,i}_0$, $i=1,\ldots ,r_h$ (= rank of $h$), in the respective sectors.
These generators may also be diagonalized. The sum, $J^{tot,i}_0$, is a BRST
 exact operator (cf. eq.(\ref{jtot})) and, for non-trivial BRST invariant
 states, we can again restrict to states with zero eigenvalues. We write the
 BRST charge \be
Q=\hat{Q}+M_ib^i_0+c_{0,i}J^{tot,i}_0\label{brstdecomp}.
\ee
The operator $\hat{Q}$ is nilpotent on any state for which the condition
 $J^{tot,i}_{0}\ket{s}=0$
is met. It is convenient to proceed by studying the cohomology
of $\hat{Q}$ on the relative space \cite{FGZ}
\be
b_{0,i}|s\rangle=0\hs{20mm}i=1,\ldots ,r_h.\label{relspace}
\ee
We will now prove the following result for the relative cohomology.\vs{5mm}

\begin{quotation}{\it Let $\tilde{R}$ be a representation of highest weight
 $\tilde{\mu}$ such that all states $| s_{\widetilde{H}}\rangle$ in
 eq.(\ref{state}) belong to
$\cStH$, i.e. there are no null-states in this sector, then the relative
 cohomology is non-trivial only for states $\ket{\phi}$ which have zero ghost
 number, have no $\jt^a_{-n}$-excitations and satisfy
\be
J^a_n\ket{\phi}=b^a_n\ket{\phi}=c^a_n\ket{\phi}=0,\label{brstgko}
\ee
for $n>0 \hs{2mm} {\rm \it{or},}\hs{2mm} n=0\ {\rm \it{and}}\
 a\in\{\alpha^+\}$. In addition,
\be
\mu^i+\tilde{\mu}^i+\rho^i =0,\label{hweq}
\ee
where $\rho^i=\sum_{\alpha>0}\alpha^i$ and $\mu^i$ is the weight of
 $\ket{\phi}$ w.r.t. $h$.}
\end{quotation}\vs{5mm}

It should be remarked that an equivalent statement may be made for lowest
 weight representations $\tilde{R}$. Let us now prove the results above. Note
first that eq.(\ref{hweq}) follows from eq.(\ref{brstgko}) and the condition
 $J^{tot,i}_0\ket{\phi}=0$. It implies that
the Casimirs of the representations of $h$ and $\tilde{h}$ satisfy
 $C-\tilde{C}=0$. Furthermore, the equations $J^a_n\ket{\phi}=0$ are
implied by the BRST invariance of states with no ghost-excitations.

Introduce a gradation of states using the general form eq.(\ref{state}).
 First we take $\ket{\tilde{0};\tilde{R}}$, $\ket{0^+}_{gh}\equiv\prod_
{\alpha>0}c_0^\alpha\ket{0}_{gh}$ and an arbitrary state in the $G$-sector to
 have zero degrees. Then a state is decomposed into states of definite
 degrees, which are determined by the operators acting on the ground-state
 \cite{HM}
\be
grad(\jt_{-n}^a)&=&1, \hs{8mm} {\rm  for~} n>0~~ {\rm or}~n=0,\
 a\in\{\alpha^-\}\nonumber \\
grad(b_{-n}^a)&=&1, \hs{8mm} {\rm  for~} n>0~~ {\rm or}~n=0,\
 a\in\{\alpha^-\}\nonumber\\
grad(c_{-n}^a)&=&-1  \hs{7mm} {\rm  for~} n>0~~ {\rm or}~n=0,\
 a\in\{\alpha^-\}\label{grad}
\ee
All other operators have zero degree. We note that a state with a ghost
 number $N_{gh}$ will always decompose into states of degrees that are
 greater than or equal to $-N_{gh}$. We have here taken the ghost number of
 $\ket{0^+}_{gh}$ to be zero, a convention which we will adopt throughout
 this and the next section. The gradation above is not conserved by the
 commutators. This means that the degree will depend on the ordering of the
 operators which build up the states. It is, therefore, convenient to refer
 only to the {\it maximum} degree $N$ of a state, {\it i.e.} a state which
 has a leading term of degree $N$. One has the following decomposition
 (cf. eq.(\ref{qcw})): $\hat{Q}=d_0+d_{-1}$, where
\be
d_0=\sum_{m>0} \jt_{-m}^ac_{m,a}+\sum_{\al\in\{\al^+\}}\jt_0^{-\al}c_0^{\al}
\ee
and $d_{-1}$ is the remainder. The index indicates the degree of the operator,
 which when acting on a state of a maximum degree $N$ will give a state of
 maximum degree not exceeding the sum of the degree of the operator and $N$.

We now solve the BRST equation, which in the relative space (\ref{relspace})
 implies
\be
\hat{Q}|\phi\rangle=0.\label{relbrsteq}
\ee
Let $|\phi\rangle$ be a state of maximum degree $N>0$, {\it i.e.}
 $|\phi\rangle=|s;N\rangle+\ldots$\ . Then eq.(\ref{relbrsteq}) implies to
 highest order
\be
d_0|s;N\rangle=0\label{deq}
\ee
In addition, $(d_0)^2=0+{\cal{O}}(-1)$, so that to leading order we should
 determine the cohomology of $d_0$.

Consider a set of monomials
\be
\ket{p,q}=\jt_{-n_1}^{a_1}\ldots \jt_{-n_p}^{a_p}b_{-m_1}^{e_1}\ldots
 b_{-m_q}^{e_q}|s_G\rangle|\tilde{0} ;\tilde{R}\rangle|\phi_{gh}\rangle.
\label{basis}
\ee
This set provides a basis for the full relative state space, provided the
 states $|s_G\rangle\in \cSG$ and $|\phi_{gh}\rangle$ are chosen
 appropriately, and one defines a specific ordering among the modes $\jt_m^a$.
 Let us introduce a homotopy operation on this space, defined by its action
 on the monomials \cite{HM}
\be
&&\ka_{0}|p,q\rangle\equiv
\frac{1}{p+q}\sum_{i=1}^p\jt_{-n_1}^{a_1}\ldots\widehat{\jt_{-n_i}^{a_i}}\ldots
\jt_{-n_p}^{a_p}b_{-n_i}^{a_i}b_{-n_1}^{e_1}\ldots b_{-n_q}^{e_q}\nonumber\\
&&\hs{25mm}\times|s_G\rangle|\tilde{0} ;\tilde{R}\rangle|\phi_{gh}\rangle
\hs{20mm}p\neq 0\nonumber\\
&&\ka_{0}|p=0,q\rangle\equiv 0 \label{kappa}
\ee
where the capped terms are omitted. It is not essential to our argument
 whether $\ka_0$ exists as an operator or not. However, it is easy to see
 that it does, in fact, exist. Let $\ket{s_0}$ be a state and $\ket{s_1}
\equiv\ka_0\ket{s_0}$, as defined by eq.(\ref{kappa}). If $\ket{s_0^\prime}$
 is a state for which $\langle s_0^\prime |s_0\rangle=1$, then we can realize
 $\ka_0$ as the operator $|s_1\rangle\hs{-1mm}\langle s_0^\prime|$.

Using the definition above, it is straightforward to verify the relation
\be
(\ka_{0}d_0+d_0\ka_{0})\ket{p,q}=\delta_{p+q,0}\ket{p,q}\label{kd},
\ee
which is valid to highest order.
On the state $\ket{s;N}$ this and eq.(\ref{deq}) implies
\be
\ket{s;N}=d_0\ket{s^\prime;N}+{\cal{O}}(N-1).
\ee
This in turn implies
\be
\ket{\phi}=
d_0\ket{s^\prime;N}+{\cal{O}}(N-1)=\hat{Q}\ket{s^\prime;N}+{\cal{O}}(N-1).
\ee
We have, therefore, that $\ket{\phi}$ is $\hat{Q}$-exact to highest order in
 our gradation. One proceeds in a standard fashion, concluding to each highest
 order the exactness of the state. In this way any BRST invariant state in the
 relative space is shown to be cohomologically equivalent to a state with zero
 or negative maximum degree. A state with negative ghost number will, however,
 always decompose into states of positive degrees, and therefore, will be BRST
 trivial. This implies that the states with positive ghost number are trivial
 as well, which follows from the theorem by Kugo and Ojima \cite{KO}.

We have finally only states of zero ghost number left to consider.
These states may be decomposed into states of degrees greater than or equal to
 zero. Let us assume that it has a maximal degree larger than zero. We denote
 the highest order term of $\ket{\phi}$ by
$\ket{p,N}$, where the degree $N>0$. $\ket{p,N}$ must then have some
 $\jt_{-n}^a$-excitations. This implies that in analyzing the BRST equation
 one may use the homotopy operation defined in eq.(\ref{kappa}) to conclude
 that $\ket{p,N}$ and hence $\ket{\phi}$ is BRST exact to this order. We can
 proceed in this fashion as long as the highest order term has a degree which
 exceeds zero. In this way one eliminates all $\jt^a_{-n}$-excitations. Now,
 if the state contains any $b^a_{-n}$-excitations, $\ket{s}=b^a_{-n}\ket{s^
\prime}+\{terms\  with\ no\ b^a_{-n}-dependence\}$, then by applying the BRST
 charge we get
the highest order term $\jt^a_{-n}\ket{s^\prime}+\{terms\ with\ no\ \jt^a_{-n}
-dependence\}=0$. This cannot be solved unless $\ket{s^\prime}=0$.
 We can conclude, therefore, that $\ket{\phi}$ does not contain any $b_{-n}^a$
-excitations and, hence, no ghost dependence at all, and in addition,
no $\jt^a_{-n}$-excitations.
\vs{5mm}

This concludes our proof. Before discussing the
relevance of the analysis to the coset model, we should also address the
 absolute cohomology. It is clear that the absolute cohomology contains a lot
 of more states using arguments due to \cite{FGZ},\cite{BMP1}. This follows
 from the fact that the ghost vacua has a $2^{r_H}$ degeneracy. In order to
 remove this degeneracy, we will in the next section impose that physical
 states are in the relative space.

The generalization to the cases where $H$ (and $G$)
 are not simple, but of the form
$H_1\times H_2\times\ldots$, where $H_i$ are simple or in the center of $H$,
is straightforward. Then the algebra $\hat{h}$ and $\hat{\tilde{h}}$ is a sum
 $\hat{h}_1\oplus\hat{h}_2\oplus\ldots$ and, therefore, the BRST charge
 decomposes correspondingly as $Q_1+Q_2+\ldots$. Each separate term $Q_i$ may
 then be
analyzed as above.

\section{Physical states and characters}
\setcounter{equation}{0}
\markboth{\underline{\hspace{53mm} \  \bf{Section} \thesection \ \ Physical
 states and characters }}
{\underline{\bf{Section} \thesection \ \ Physical states and characters
 \hspace{53mm} \  }}

We will now use the analysis of the preceding section to investigate the
space of physical states relevant for the coset model. We will first
remove the degeneracy due to the ghost zero modes corresponding to the
Cartan subalgebra. We impose, therefore, the additional conditions
\be
b^i_0\ket{{\rm phys}}=0\hs{20mm}\mbox{for } i=1,\ldots ,r_H\label{relspacei}
\ee
This implies that we only need to consider the relative cohomology. According
 to the results of the relative cohomology
there exists a {\it unique} highest weight solution which is of zero ghost
 number (relative to the state $\ket{0^+}$) provided there are no null-states
 w.r.t. $\hat{\tilde{h}}$. It satisfies
\be
J^a_n\ket{{\rm phys}}=\jt_n^a\ket{{\rm phys}}=b^a_n\ket{{\rm phys}}=c^a_n
\ket{{\rm phys}}=0,\label{physi}
\ee
for $n>0$ or $n=0$, $a\in\{\alpha^+\}$. These conditions are equivalent to
the conventional coset conditions (\ref{coset}). In addition, there exists a
 corresponding lowest weight solution. One must now address the question of
 what representations of $\tilde{h}$ do not have null-states. We will
 establish the following important conclusion:
\begin{quotation}
{\it For integrable representations of $g$, and for representations of
$\tilde{h}$ satisfying eq.(\ref{hweq}), the only solution to the BRST
equation in the relative space are states satisfying eq.(\ref{physi}).}
\end{quotation}
A different way of phrasing this result is as follows: If we only take
 representations
of $\tilde{h}$, such that the states satisfying the coset
conditions (\ref{physi}) are at least contained in the set of solutions
 of the BRST equation, then these states are, in fact, the only possible
 solutions.

To prove this statement, it is sufficient to prove that, for representations
satisfying eq.(\ref{hweq}), there are no null-states w.r.t. $\hat{\tilde{h}}$
if we restrict to integrable representations of $g$. Let us, therefore,
 investigate the null-states w.r.t.
$\hat{\tilde{h}}$. This may be done by  examining the Kac-Kazhdan determinant
\cite{KK}. We first consider the simplest case
$\hat{su}(2)$. Then the null-states are parametrized by two integers $n$ and
 $n^\prime$.
If the highest weight of the ground-state is denoted by $j$ ($2j\in {\cal Z}$),
then the highest weight of the primary null-state is $j+n$ and the state occurs
at occupation level $N=nn^\prime$ for the representations \be
2j+1=-n+n^\prime(k+2).\label{nullrepr}
\ee
Here $n,n^\prime\geq 1$ or $n\leq-1, n^\prime\leq 0$. For $\hat{\tilde{h}}$ we
 have
$\tilde{k}=-k-4$, so that for the highest weight representations $\tilde{j}$
 in this sector we have
\be
2\tilde{j}+1=-n-n^\prime(k+2),\label{nullreprt}
\ee
with $n$ and $n^\prime$ as before. We see that for
$-k-2\leq 2\tilde{j}+1\leq 0$ there are no null-states. Then by
eq.(\ref{hweq}) we have $\tilde{j}=-j-1$, which
implies that $-1/2\leq j\leq (k+1)/2$. This range includes all the integrable
representations of $su(2)$, $0\leq j\leq k/2$. Thus, for the integrable
representations of $h=su(2)$ there are no null-states for $\hat{\tilde{h}}$
 and,
therefore, the coset conditions eq.(\ref{physi}) are the unique solutions. We
note that apart from the values $j=-1/2$ and $(k+1)/2$, the representations
outside
the range of integrable ones will not contain the usual coset conditions. The
representations $-k-2\leq 2\tilde{j}+1\leq 0$, having negative highest weights,
are infinite dimensional discrete representations and correspond to unitary
representations of $SU(1,1)$.

We now consider general simple algebras $\hat{h}$.
Let $\mu^i$ denote the highest weight of the ground-state, then the
primary null-state has highest weight $\mu^i+n\alpha^i$, where $n$ is an
 integer
and $\alpha^i$ a positive root, and occur at occupation level $N=nn^\prime$ for
representations
\be
(2\mu +\rho)\cdot\alpha=-n\alpha^2+n^\prime(k+c_H)\label{gzeros},
\ee
where $n$,$n^\prime\geq 1$ or $n\leq -1$,$n^\prime\leq 0$. From this
 expression we deduce that it is sufficient that $(1/\alpha^2)(2\mu +\rho)
\cdot\alpha \leq
0$ for all positive roots and
\be
2\hat{\alpha}\cdot\tilde{\mu}+1\geq \tilde{k}+2,\label{nonull}
\ee
for
$\hat{\alpha}^i$ being the highest root, to have
have no null-states. Thus, for these
representations we will only have the physical states satisfying the coset
conditions  (\ref{physi}). Then eqs.(\ref{hweq}) and (\ref{nonull}), using
that $\hat{\alpha}\cdot \rho=c_H-1$ and $\tilde{k}=-k-2c_H$, imply that
$(1-c_H)/2\leq \hat{\alpha}\cdot\mu\leq
(k+1)/2$. This again includes all integrable
representations, $0\leq\hat{\alpha}\cdot\mu\leq k/2$, which, consequently,
proves the uniqueness of the coset conditions for arbitrary groups.

We now turn to the characters of the gauged WZNW models. Having
proved the uniqueness of the BRST invariant physical states, we
are assured that if we construct the characters in such a way that
the BRST symmetry is respected, then only the correct physical
degrees of freedom will propagate. Generally the character is
defined as (we will omit the conventional factors of $e^{-2i\pi\tau c/24}$)
 \be
\chi(\tau,\theta)=Tr\left(e^{2i\pi\tau (L_0^G+L_0^{\ti{H}}+L_0^{gh})}
e^{i(\theta_iJ^{tot,i}_0+\theta_{I^{\prime}}J_0^{I^{\prime}}})
(-1)^{N_{gh}}\right),\label{char}
\ee
where $ I^{\prime}\in g/h$\footnote{We assume here that $h$ is embedded in $g$
 in such a way that this decomposition is possible.}  In taking
the trace, the projection onto the relative space, eq.(\ref{relspacei}), must
be implemented. Therefore, the trace does not include a summation over the
corresponding ghost vacua. This is consistent with the BRST symmetry only if
we, in addition, require that the commutator of the conditions
(\ref{relspacei}) with the BRST charge vanishes, {\it i.e.} a projection onto
states satisfying $J^{tot,i}_0\ket{s}=0$ is made. We define,
consequently, the BRST invariant character of the $G/H$ WZNW model as
\be
\chi^{G/H}=\int\prod_i\frac{d\theta_i}{2\pi}Tr\left(e^{2i\pi\tau(L_0^G+
L_0^{\ti{H}}+L_0^{gh})}e^{i(\theta_iJ_0^{tot,i}+\theta_{I^{\prime}}
J_0^{I^{\prime}})}(-1)^{N_{gh}}\right).\label{chari}
\ee
The trace here contains no summation over ghost zero modes $c^i_0$ and
$b^i_0$ and the integration over $\theta_i$ projects onto states
satisfying $J^{tot,i}_0\ket{s}=0$.

The characters (\ref{chari}) is a product of three different terms $\chi^G$,
$\chi^{\tilde{H}}$ and $\chi^{gh}$ due to the corresponding factorization of
the state space. The first factor is the character of an arbitrary $G$ WZNW
model and for a simple group is given by the Kac-Weyl formula \cite{KACII}.
The second factor is the character of $\hat{\tilde{h}}$. This character is
straightforward to determine, since the state space is free of null-states
for the representations we have selected. If the highest weight of the
representation is denoted by $\tilde{\mu}$, then the character is
given by
\be
\chi^{\tilde{H}^+}(\tau,\theta) =
e^{i\theta\cdot\tilde{\mu}}e^{(-\frac{2i\pi\tau }{k+c_H}(
\tilde{\mu}\cdot(\tilde{\mu}+\rho))}R^{-1}(\tau,\theta)\label{chart}.
\ee
Here $\rho$ is the sum of positive roots and \be
R(\tau,\theta)=\prod_{n=1}^\infty\left[ (1-e^{2i\pi n\tau})^{r_H}
\prod_{\alpha>0}(1-e^{2i\pi (n-1)\tau}e^{-i\alpha\cdot\theta})(1-e^{2i\pi
 n\tau}
e^{i\alpha\cdot\theta})\right].\label{delta}
\ee
The corresponding character for a representation with the
lowest weight $(-\tilde{\mu})$ is given by
\be
\chi^{\tilde{H}^-}(\tau,\theta)
= -e^{(-i\theta\cdot(\tilde{\mu}+\rho))}e^{(-\frac{2i\pi\tau }{k+c_H}
 \tilde{\mu}\cdot
(\tilde{\mu}+\rho))}R^{-1}(\tau,\theta).
\ee
The character of the ghost sector is
also easily found. For each ghost pair $b^a_n$ and $c^a_n$ we find the same
contribution as for the conformal ghosts, with the exception that it is twisted
according to the eigenvalue of $J^{gh,i}_0$. The trace over zero modes
$b^\alpha_0$ and $c^\alpha_0$ yields
$e^{i\rho\cdot\theta}\prod_{\alpha>0}(1-e^{-i\alpha\theta})^{2}$.
Consequently, we have
\be
\chi^{gh}(\tau,\theta)=e^{i\rho\cdot\theta}R^2(\tau,\theta).\label{ghchar}
\ee

We will now consider some explicit examples to see that the characters above
combine to yield results previously derived by other methods. First we take
the simplest case of an Abelian group $H$, the parafermion theory
$SU(2)_k/U(1)$. From the definition of the character of the coset
eq.(\ref{chari}) we have ($q\equiv
e^{2i\pi\tau}$)
\be
\chi^{SU(2)/U(1)}(\tau)=\int_{-\infty}^\infty \frac{d\theta}{2\pi}
 \hs{2mm}Tr\left(
q^{(L_0^{SU(2)}+L_0^{\tilde{H}}+L_0^{gh})}e^{i\theta(J^3_0+\tilde{J}^3_0)}
\right).\label{chardef}
\ee
Here $L_0^{\tilde{H}}=-(\jt_0^3)^2/k$. The trace over excited modes of
$\tilde{U}(1)$ gives $\prod_{n=1}^{\infty}(1-q^n)^{-1}$ and by integrating over
$\theta$ we will impose $J^3_0+\tilde{J}^3_0=0$. The ghost contribution is
$\prod_{n=1}^{\infty}(1-q^n)^2$. Consequently, we are left with a trace over
$SU(2)$-sector of the form
\be
\chi^{SU(2)/U(1)}(\tau)=Tr\left(
q^{(L_0^{SU(2)}-\frac{1}{k}(J_0^3)^2)}\right)\prod_{n=1}^{\infty}(1-q^n)
\label{charii}
\ee
This expression gives the branching function of $\hat{su}(2)$ w.r.t.
 $\hat{u}(1)$ and is, up to a factor of
$q^{c^{(PF)}/24}\prod_{n=1}^{\infty}(1-q^n)$, a sum of
string-functions. We may also obtain an explicit expression
for the string-functions by inserting the expressions for the character of the
$SU(2)$ and $U(1)$ theories before performing the integration over $\theta$.
The character for $SU(2)_k$ is given by
\be
\chi_{j,k}(q,\theta)=\Delta_{k,j}(q,\theta)R^{-1}(q,\theta),\label{suchar}
\ee
where
\be
\Delta_{k,j}(q,\theta)=
q^{{j(j+1)\over k+2}}\sum_{n\in{\cal Z}}
q^{(k+2)n^2+(2j+1)n}(e^{i(j+(k+2)n)\theta}-e^{-i(j+1+(k+2)n)\theta}).
\label{dkj}
\ee and
\be
R(q,\theta)=(1-e^{-\theta})\prod_{n=1}^\infty (1-q^n)(1-q^ne^{i\theta})(1-q^n
e^{-i\theta})\ee
The $U(1)$
theory is equivalent to a free boson compactified on a radius $\sqrt{k/2}$. The
character is  \be
\chi^{\tilde{H}}(q,\theta)=\sum_{m\in{\cal Z}/2}q^{-\frac{\tilde{m}^2}{k}}
e^{i\theta\tilde{m}}\prod_{n=1}^\infty (1-q^n)^{-1},\label{uchar}
\ee
so that
\be
\chi^{SU(2)/U(1)}(\tau)=\int_{-\infty}^\infty \frac{d\theta}{2\pi}
 \sum_{\tilde{m}\in
{\cal Z}/2}\Delta_{k,j}q^{\frac{-\tilde{m}^2}{
k}}e^{i\theta\tilde{m}}R^{-1}(\tau,\theta)\prod_{n=1}^\infty(1-q^n).
\ee
Expanding $R^{-1}(\tau,\theta)$ \cite{THORN},\cite{HUI},
\be
R^{-1}(\tau,\theta)=\prod_{n=1}^\infty(1-q^n)^{-3}\sum_{p=-\infty}^{\infty}
\sum_{s=0}^{\infty}(-1)^sq^{[\frac{1}{2}(s-p+1/2)^2-\frac{1}{2}(p-1/2)^2]}
e^{ip\theta}
\ee
and integrating over $\theta$ gives the final result
\be
\chi^{SU(2)/U(1)}(\tau) = q^{j(j+1)\over
k+2}\prod_{n=1}^\infty(1-q^n)^{-2}
\sum_{p,n\in {\cal Z}}\sum_{s=0}^\infty
(-1)^sq^{(k+2)n^2+(2j+1)n}\nonumber\\
\times  q^{\frac{1}{2}(s-p+1/2)^2-\frac{1}{2}(p-1/2)^2}
\left(q^{-\frac{1}{k}(p+j+(k+2)n)^2}-q^{-\frac{1}{k}(p-j-1-(k+2)n)^2}\right).
\ee
This expression for the $SU(2)/U(1)$ parafermion theory was first derived
in \cite{DISQUI} using a free field realization.

Our next example is $SU(2)_k\times SU(2)_1/SU(2)_{k+1}$,
which is the
unitary series of the minimal models. We choose the Cartan subalgebra of $G$ to
be spanned by $J^{3(1)}_0+J^{3(2)}_0$ and $J^{3(2)}_0$, where the superindices
 $(1) $ and
$(2)$ refer to the level $k$ and level one $SU(2)$ theories, respectively. We
will, for simplicity, suppress the dependence on $J^{3(2)}_0$ and define the
character as
\be
\chi^{Vir}_{j_1,j_2,\tilde{j}}=\int_{-\infty}^\infty \frac{d\theta}{2\pi}
 \hs{2mm}Tr\left(
q^{L_0}e^{i\theta J^3_0}\right)
\ee
where $L_0$ and $J_0^3$ contain the sum over all sectors. The character for the
$SU(2)_k$ theory is given by eq.(\ref{suchar}). For $k=1$ it simplifies to
\be
\chi_{j_2,1}=\sum_{\lambda \in {\cal
Z}+j_2}q^{\lambda^2}e^{i\lambda\theta}\prod_{n=1}^\infty (1-q^n)^{-1},
\ee
where $j_2=0$ or $1/2$. The character for the auxiliary $SU(2)_{-k-5}$
theory is found from eq.(\ref{chart}),
\be
\chi_{\tilde{j},-k-5}=q^{-{\tilde{j}(\tilde{j}+1)\over k+3}}e^{i\theta
\tilde{j}}R^{-1}(\tau,\theta), \ee
and the ghost contribution from eq.(\ref{ghchar}),
$\chi^{gh}=e^{i\theta}R^{2}(\tau,\theta)$. Thus,
\be
\chi^{Vir}_{j_1,j_2,\tilde{j}}=\int_{-\infty}^\infty \frac{d\theta}{2\pi}
\prod_{n=1}^\infty
(1-q^n)^{-1}\Delta_{k,j_1}(q,\theta)q^{-{\tilde{j}(\tilde{j}+1)\over k+3}}
e^{i\theta (\tilde{j}+1)}\sum_{\lambda\in {\cal
Z}+j_2}q^{\lambda^2}e^{i\lambda\theta} \ee
In performing the integration over $\theta$, we will get delta-functions,
$\delta_{\lambda +j_1+\tilde{j}+1+(k+2)n,0}$ and $\delta_{\lambda
 +\tilde{j}-j_1-(k+2)n,0}$, from the two factors in eq.(\ref{dkj}). They are
 zero unless
$j_1-\tilde{j}=j_2$ mod $1$. We can use the delta-functions to eliminate the
 sum over
$\lambda$, so that
\be
\chi^{Vir}_{j_1,j_2,\tilde{j}}&=&\prod_{n=1}^\infty
(1-q^n)^{-1}q^{j_1(j_1+1)\over k+2}q^{-{\tilde{j}(\tilde{j}+1)\over k+3}}
\sum_{n\in {\cal
Z}}q^{(k+2)n^2+(2j_1+1)n}\nonumber\\
&\times & \left[q^{(j_1+(k+2)n+\tilde{j}+1)^2}-
q^{(j_1+(k+2)n-\tilde{j})^2}\right]
\ee
If we set $p=2j_1+1$ and $r=-(2\tilde{j}+1)$, then for integrable
representations, $1\leq p\leq m-1$ and $1\leq r\leq m$, $(m\equiv k+2)$.
The range of $r$ is here determined by $\tilde{j} =-j-1$ and $j$ being an
integrable representation of $SU(2)_{k+1}$. Furthermore, from
$j_1-\tilde{j}=j_2$ mod $1$, we have $p-r$ is even or odd if
$j_2$ is $0$ or $1/2$, respectively. The character may then be written as
\be
\chi^{Vir}_{(p,r)}=\prod_{s=1}^\infty
(1-q^s)^{-1}\sum_{n\in {\cal Z}}\left( q^{\alpha_{p,r}^m(n)}
 -q^{\beta_{p,r}^m(n)}\right)\label{charvir}
\ee
with
\be
\alpha_{p,r}^m(n)&=&{[2m(m+1)n-rm+p(m+1)]^2-1\over 4m(m+1)}\nonumber\\
\beta_{p,r}^m(n)&=&{[2m(m+1)n+rm+p(m+1)]^2-1\over 4m(m+1)}
\ee
This expression is identical to the one derived by Goddard, Kent and Olive
\cite{GKO}, obtained by factorizing the $SU(2)_k\times SU(2)_1$
characters, and to the one given by Rocha-Caridi \cite{ROCHA} for the discrete
series. The range of $r$ given here is not the same as in ref.\cite{GKO}, but
in summing over $p$ and $r$, it may be extended to the same range by using the
symmetry properties of the characters. We see from the derivation above that we
never needed to use the factorization property of $SU(2)_k\times
SU(2)_1$ as a product $SU(2)_{k+1}\times Vir$. Finally, if we would have
used the lowest weight representations of the auxiliary $SU(2)$ theory, we
would have arrived at the same expression (\ref{charvir}).\vs{5mm}

Let us end by briefly discussing correlation functions. We have selected a
 particular range of representations for the auxiliary theory. We must,
 therefore, address the important issue whether this selection is consistent
in correlation functions or if the fusion rules require that we consider a
 larger set of representations. If we have two vertex operators corresponding
to the highest weights $\tilde{\mu_1}$ and $\tilde{\mu_2}$, then
these can only fuse to a vertex of highest weight starting at
 $\tilde{\mu_1}+\tilde{\mu_2}$ and continuing downwards to lower weights. This
follows from the conservation of the eigenvalues of the Cartan generators of
 the finite algebra. In order for our selection of representations to be
 consistent,
there must exist a consistent truncation of this series of possible
 representations. We will now show that this is the case. We start again by
 considering
$SU(2)$. In this case we may make use of a decoupling theorem originally proved
for $SU(1,1)$ \cite{HwRo}, modifying the original theorem of Gepner and Witten
 \cite{GW}. The important point is not whether we have $SU(1,1)$
or $SU(2)$, but if the level is positive or negative. We make use of the
 highest weight null-vectors
\be
(\tilde{J}^+(z))^NV_{\tilde{j}}(z)=0.\label{nullpr}
\ee
Here $\tilde{j}$ refers to the highest weight, which is a negative integer or
 half-integer for the infinite dimensional representations which we are
 considering. These null-vectors occur for $\tilde{j}=\tilde{k}/2-(N-1)/2$ for
$N=1,2,3,\ldots $. Since $\tilde{k}$ is negative, the null-vectors are present
for all representations in the infinite dimensional discrete series, except
those for which $\tilde{j}>\tilde{k}/2$. They are also absent for the trivial
representation. Inserting (\ref{nullpr}) into a three-point function will lead
 to
restrictions on this correlation function. It will be zero whenever we mix
 vertices with null-vectors and vertices without. This is where the sign of the
level enters. For positive levels the trivial representation is of the former
type, while for negative levels it is of the latter type. Consequently, for
negative levels the vanishing of the three-point function can only imply that
representations with null-vectors decouple (or else there is no propagation at
 all) {\it i.e.} all highest weight representations for which $\tilde{j}\leq
\tilde{k}/2$
decouple. This is the decoupling theorem. Note, that the fact that the coupling
vertices do not contain null-vectors implies that we do not get further
 restrictions on the couplings, contrary to the case of positive level.

The generalization to arbitrary groups proceeds in the same way as our
 analysis of null-vectors
in the beginning of this section. One selects an $\hat{su}(2)$ subalgebra
 spanned by $\tilde{J^\alpha_m}$, $\tilde{J^{-\alpha}_m}$ and $\frac{1}
{\alpha^2}\alpha\cdot \tilde{J}_m$. Then by using the corresponding
null-vectors, one reaches the conclusion that we have decoupling unless the
highest weights are in the range $3/2-c_H\leq \alpha\cdot\mu\leq
(k+1)/2$, which is consistent with the selection of weights we have
 considered. This concludes the decoupling for the general case.

The fact that we have a consistent truncation of representations is perhaps
 not so surprising from the point of view of the original GKO construction.
 In this
formulation, the vertices of the coset have to commute with the generators
of $\hat{h}$. One may then use this to show \cite{BG} that such vertices will
 close
under fusion. In the present formulation we have seen that the conventional
 coset
 condition follows
from the BRST condition. Therefore, it is natural to expect that the closure
in the GKO construction follows from a corresponding closure in the BRST
 formulation.
\newpage

{\large{\bf{Acknowledgments}}}
\\
\noindent It is a pleasure to thank Arne Kihlberg for discussions on group
 theory. We would also
like to thank Robert Marnelius and Christian Preitschopf for discussions and
Antti Kupiainen for explaining certain issues in ref.\cite{GK}.

\end{document}